\documentstyle[prd,aps,eqsecnum,preprint,psfig,floats]{revtex}
\begin{document}
\draft

\preprint{FSU-HEP-971020}

\title{Jet photoproduction and the structure of the photon}

\author{B.W. Harris and J.F. Owens}

\address{Physics Department \\ Florida State University \\
Tallahassee, Florida 32306-4350, USA}

\date{October 1997}

\maketitle

\begin{abstract}

Various jet observables in photoproduction 
are studied and compared to data from HERA.
The feasibility of using a dijet sample for constraining the 
parton distributions in the photon is then studied.
For the current data the experimental and theoretical 
uncertainties are comparable to the variation due
to changing the photon parton distribution set.

\end{abstract}


%---------------Section 1--------------------------
\section{Introduction}
%--------------------------------------------------
Large momentum transfer processes involving photons are useful for studying 
strong interaction dynamics, for studying hadronic structure, and for refining 
our understanding of the photon-hadron interaction itself. The photoproduction
of hadronic jets is an example of such a process. 
In a previous paper \cite{ho}, a complete ${\cal O}(\alpha \alpha_s^2)$ 
calculation of jet photoproduction, based on the
phase space slicing technique, was described 
and predictions were compared with data for several single jet and dijet
observables. Since the completion of that analysis additional data have become
available. The new data have increased statistics and have been analyzed using
several different jet definitions. Results for several new distributions
involving dijets are also now available. Therefore, it is appropriate to
investigate the degree to which the new data are described using the program
described in Ref.\ \cite{ho}.

An important aspect of hard photon-hadron interactions is the existence of two
different interaction mechanisms. The ``direct'' component corresponds to the
case where the photon participates wholly in the hard scattering, giving its
full energy to the underlying photon-parton subprocesses. The ``resolved'' 
component of the photon \cite{dg} corresponds to the case where the photon
generates a shower in the initial state of nearly collinear quarks and gluons,
one of which interacts with a parton from the initial state hadron. In
this case it is convenient to define parton distributions in the photon and to
sum the resulting collinear logarithms using the Altarelli-Parisi formalism
appropriately extended to the photon case \cite{ap}.  
Information on the parton distributions in the photon is available from 
studies of the photon structure function $F_2^{\gamma}$, 
jet production in $e^+e^-$ two photon collisions, 
heavy quark production in $e^+e^-$ two photon collisions, 
and in jet photoproduction.  For recent overviews, see, for example 
Refs.\ \cite{forshaw} and \cite{aurenche}. 
For the first two processes, the lowest order diagrams involve only quarks. 
On the other hand, the resolved component of 
heavy quark production in two photon collisions \cite{hoh} 
and jet photoproduction both in lowest order include subprocesses with 
initial state gluons.  Therefore, at least in principle, they should provide
information on the gluon distribution in the photon which would complement 
that obtained from the other two processes.  Some of the issues discussed 
in this paper have been reviewed in \cite{klasen}.

The goal of this paper is to critically examine the current theoretical
description of jet photoproduction and, as a result, assess the ability of the
data to distinguish between different parametrizations of photon parton
distributions. In Sec.\ II the predictions for inclusive single jet production
are compared with the latest available data. A similar comparison for various
dijet observables is presented in Sec.\ III. 
Conclusions and suggestions for further study are given in Sec.\ IV.

%---------------Section 2--------------------------
\section{Single Jet Inclusive Distributions}
%--------------------------------------------------

If jet photoproduction is to serve as a reliable source of information 
on the parton distributions in the photon, then it is necessary to assess the
quality of the theoretical description of the data. Differences between the
theory and the data could be due to several sources, {\it e.g.,} an inadequate
description of the underlying hard scattering subprocesses or incorrect parton
distributions in the initial state photon or hadron. In order to examine these
issues, we first consider the inclusive single jet rapidity distributions. 

The predictions for all of the jet observables in this paper have been 
generated with one or more variants of the cone algorithm for jet merging 
described in Ref.\ \cite{ho}.  This algorithm depends on a parameter 
$R_{\rm sep}$ \cite{rsep} governing the merger of two partons that 
are widely separated but, none the less, lie within a cone of the 
prescribed radius $R$.  This variable lies in the range 
$R \le R_{\rm sep} \le 2R$.  For $R_{\rm sep}=2R$ the algorithm reduces 
to that of Ref.\cite{snow} and the introduction of $R_{\rm sep}$ has 
no effect.  At the other extreme, $R_{\rm sep}=R$, the algorithm applied 
to three parton final states is equivalent to the $k_T$ algorithm 
\cite{ktalg}. 

Recent studies of jet shapes \cite{shape1}, \cite{shape2} 
find good agreement between data analyzed using the iterative 
cone algorithm and theory if the problem of merging 
overlapping jets is taken into account by varying the parameter 
$R_{\rm sep}$ as a function of transverse energy and rapidity. The optimal
values of $R_{\rm sep}$ varied slowly with $E_T$ and $\eta_{\rm jet}$ and 
values in the 
range of $1.3-1.4$ were obtained for most of the region covered by the data. 

In Fig.\ 1 the data for the single-jet inclusive cross section as a function 
of the jet pseudorapidity 
$\eta_{\rm jet}$ integrated over $E_T^{\rm jet}>E_T^{\rm min}$ for 
$E_T^{\rm min} = 14,17,21$, and $25\,$GeV as measured by ZEUS \cite{zeuscone} 
using an iterative cone jet finding algorithm with $R=1$ are shown.  These 
data are compared with the results of our next-to-leading calculation using 
the Aurenche-Fontannaz-Guillet (AFG) \cite{afg} photon distributions and 
the CTEQ4M \cite{cteq4} proton parton distributions. The three curves 
correspond to $R_{\rm sep}=2$ (top), $R_{\rm sep}=1.3$ (middle), and 
$R_{\rm sep}=1$ (bottom). The factorization and renormalization scales have 
both been chosen to be $E_T^{\rm max}$ where $E_T^{\rm max}$ is the largest 
transverse jet energy in the event. 
This comparison clearly shows that the theory underestimates the
data in the region $\eta_{\rm jet} > 0.5$ for the lowest $E_T^{\rm min}$ set, 
with the discrepancy decreasing with increasing $E_T^{\rm min}$.  
This result is consistent with that presented
in \cite{ho}. One possibility for this discrepancy is that in the forward
region fragments of the proton beam jet are being included in the jet cone and,
therefore, the energy of the observed jet is increased. The effect of such a
contribution would be decreased if a smaller cone radius was used to define
the jet. Therefore, data have also been presented corresponding to a cone of
$0.7$. These are shown in Fig.\ 2 together with the corresponding theoretical
predictions. The agreement is now much improved in the forward direction.  

The use of a smaller cone size to define the jets appears to solve the problem
of the jet excess in the forward $\eta_{\rm jet}$ region.  The agreement 
between theory
and experiment shown in Fig.\ 2 increases the confidence one has in the
theoretical description of jet photoproduction.  However, the inclusive 
single-jet observables still involve integrations over unobserved jets with a
corresponding loss of information. In order
to further test the theoretical description it is necessary to look at
observables which involve more than one jet. While there will still be
integrations over unobserved jets or partons, dijet observables offer more 
ways to test the theory. 

%---------------Section 3--------------------------
\section{Dijet Observables}
%--------------------------------------------------

A fundamental requirement for utilizing jet photoproduction as a means of
constraining photon parton distributions is that the underlying hard scattering
subprocesses must be correctly described. A sensitive test of this is to study
the dijet angular distribution in the dijet center of mass
system. In lowest order, dijet production is described in terms of $2
\rightarrow 2$ subprocesses and the two jets will be back-to-back with
balancing transverse momenta. The angular distribution of the dijet axis with
respect to the beam is a fundamental prediction of the theory and its 
shape is only slightly modified when higher order effects are taken into 
account \cite{ho}.

The top portion of Fig.\ 3 shows the inclusive dijet cross section as a 
function of the absolute value of the cosine of the angle between 
the dijet axis and the 
beam axis $ | \cos\theta^* | $ with the dijet mass $M_{\rm JJ}$ greater 
than $47\,$GeV.  The data were measured by ZEUS \cite{zeusdijet} using 
their iterative cone algorithm with $R=1$ and are compared with our 
next-to-leading order result for $R=1$ with $R_{\rm sep}=2R$ (solid) 
and $R_{\rm sep}=R$ (dot).  This particular 
observable was found to be relatively insensitive to the choice of the photon 
parton distribution set. The predictions shown here were obtained using the 
Gordon-Storrow (GS) distributions \cite{gs}.
The overall agreement is good, with the trend of the data being between the
curves for the two values of $R_{\rm sep}$. The good description of the angular
distribution data suggests that the underlying subprocesses are correctly
described. 

Next, consider the dependence of the cross section on the mass of the dijet
system.  In lowest order the dijet mass is given by 
$M_{\rm JJ}=\sqrt{x_{\gamma} x_p s}$ where $x_{\gamma}$ and $x_p$ are the
momentum fractions carried by the partons from the photon and proton, 
respectively, and $s$ is
the square of the photon-proton center of mass energy. Hence, the dijet mass
distribution will be sensitive to the shapes of the parton distributions. 
Since $\cos\theta^*$ is being integrated over, the shape 
of the dijet mass distribution does not depend too sensitively on the 
shape of the angular distribution. 

The lower part of Fig.\ 3 shows the same data as in the upper part, but
integrated over $\cos\theta^*$ and displayed versus $M_{\rm JJ}$.  
Again, the agreement is good, lending support for the description 
of the overall production process.

We next consider the $E_T$ dependence of the highest $E_T$ jet in dijet events
with the pseudorapidities of the two jets constrained to be in specified 
ranges. 
Data measured by ZEUS \cite{zeusdijet} are shown in Fig.\ 4 together with 
our theoretical results generated with the GS photon distributions and with
both the renormalization and factorization scales equal to $E_T^{\rm max}$. 
The events are symmetrized in $\eta_1$ and $\eta_2$ so there are two 
entries per event. Furthermore, the highest $E_T$ jet
is required to have $E_{T_1} > 14\, {\rm GeV}$ while the second jet can have an
$E_T$ as low as $11$ GeV. This asymmetric $E_T$ requirement avoids a 
potential problem with the definition of the dijet sample \cite{ho,ggk,kk2}. 
The lower 
curve of each pair has an additional cut of $x_{\gamma}>0.75$ placed on it, 
where $x_{\gamma}$ is the momentum fraction of the parton from the photon 
and is reconstructed using \cite{xgamdef}
\begin{equation}
x_{\gamma}=( E_{T_1} e^{-\eta_1}
          +  E_{T_2} e^{-\eta_2} )/2E_{\gamma}
\end{equation}
with $E_{\gamma}$ the photon energy in the lab frame.
This cut removes much of the resolved component, so the lower curve of each 
pair is dominated by the direct contribution. The jet definition used for these
data is the $k_{\perp}$ algorithm \cite{ktalg}.   It has several 
advantages over the cone algorithm, although an extensive discussion of 
these points is beyond the scope of this work.  The interested reader 
is pointed towards the original literature \cite{ktalg} and recent 
detailed studies \cite{seymour}, \cite{workj} for details. 
For the three parton final 
states considered in this calculation, the $k_{\perp}$ algorithm corresponds 
to setting $R_{\rm sep}=R=1$.

Over most of the $\eta_1$ and $\eta_2$ ranges shown in Fig.\ 4, the
comparison between the theoretical results and the data appears to be good.
The exception occurs when one or both of the jets has a negative
pseudorapidity, with the discrepancy being worst if both jets have negative 
pseudorapidities. This point will be discussed in more detail below.

These same data are shown in Fig.\ 5 versus $\eta_2$ in bins of $\eta_1$ and
integrated over $E_T > 14\, {\rm GeV}$. For each set of data, curves for three
choices of the common renormalization and factorization scales are shown: 
$\mu = E_T^{\rm max}/2, E_T^{\rm max}$, and $2E_T^{\rm max}$ all using the 
GS photon set.  In Fig.\ 6 these data 
are compared to the theoretical results obtained using the AFG \cite{afg}, 
GS \cite{gs}, and Gl\"{u}ck-Reya-Vogt (GRV) \cite{grv} photon parton 
distributions all using $\mu = E_T^{\rm max}$. From the results
shown in Figs.\ 5 and 6, it can be seen that the variations due to the scale
choice are comparable in magnitude to those resulting from the different 
choices of photon parton distributions.  
Furthermore, these variations are comparable to the
experimental uncertainties on the data.  However, it should be noted that 
the variation due to the change of scale leads to an overall shift in the 
normalization without an appreciable change in shape.  Hence, even allowing 
for this scale uncertainty, constraints on the photon parton distributions 
are provided by the shapes of such observables.  
Generally speaking, the data are well
described within the theoretical and experimental uncertainties. However, a
clear discrepancy is apparent in the region of negative $\eta_1$ and $\eta_2$ 
(the left-most data point in the center and left-hand plots in these two
figures).

In order to further examine this disagreement, the data are shown again in 
Fig.\ 7 and Fig.\ 8, the latter having the cut $x_{\gamma} > 0.75$ imposed. 
In each case, four curves are shown corresponding to the full result 
(solid line), the resolved component with only quarks in the photon (medium 
dashed line), the resolved component with only gluons in the photon (dotted 
line), and direct component (long dashed line).  
Several points are immediately clear.  
The negative pseudorapidity region is dominated by the
direct component.  Hence, no amount of variation of the photon parton
distributions will bring the data and theory in line.  
Second, the gluon contribution is small nearly 
everywhere for the kinematics of the current data sample. Only for both
rapidities in the forward direction (right-most region of the right-hand plots)
is the gluon distribution in the photon making a significant contribution.

The HERWIG \cite{herwig} Monte Carlo, which uses the parton-shower 
approach for intial-state and final-state QCD radiation including 
colour coherence and azimuthal correlations, both within and between jets,
was also compared with the data. 
The results are shown in Fig.\ 9.  The HERWIG results have been scaled
upwards by a factor of 1.6, but the shape is in good agreement with
the data. Next, Fig.\ 10 shows the ratio of the HERWIG results to the 
HERWIG results at the leading-order matrix element level, {\it i.e.}, 
treating the two scattered partons as being the jets with 
all showering and subsequent hadronization off.  It can be
seen that there is a correction which corresponds to a depletion of events in
the negative rapidity region. This represents the effects of the showering and
subsequent jet reconstruction slightly shifting the energies of the jets. Near
the edge of the kinematic region covered by the data this results in a decrease
in the cross section.  This decrease makes the full HERWIG results
steeper in this region and gives a better description of the data. Hence, it
appears that the origin of the inability of the next-to-leading-order
calculation to describe this region properly is that the single parton
branching that occurs in a three-body final state does not give a sufficient
description of the full showering.

An estimate of the amount of showering correction coming from the
next-to-leading-order terms can be obtained by looking at the ratio of the
results of our calculation using the full next-to-leading-order 
matrix elements 
to those obtained using the leading-order matrix elements only. This ratio 
shows to what extent the showering correction is modeled by the three-body
final states and is shown in Fig.\ 11.  In the
region where the discrepancy exists, the ratio is relatively flat. This shows
that the shapes of the leading-order and next-to-leading-order results are 
very similar and this, in turn, implies that the next-to-leading-order
calculation will not give a correct description of the data in that region.
Note that as $\eta_2$ is further decreased the ratio rapidly increases.  
This phenomenon has been anticipated in dijet production at the Tevatron
\cite{edge} and is due to a kinematic 
suppression of the two-body contribution relative to the three-body 
contribution near the edge of phase space.  Note, however, that this effect 
occurs {\em outside} the region covered by the data in this case.

The net result is that one must include a showering correction in
order to describe the data points with the most negative pseudorapidity 
values. On the other hand, this region is dominated by the direct component 
and, therefore, is not critical for the purpose of constraining the parton
distributions in the photon.

%---------------Section 4--------------------------
\section{Conclusion}
%--------------------------------------------------

As new data have accumulated, the level of understanding of jet photoproduction
has increased. When a cone size of $R=0.7$ is used to define the experimentally
measured jets, the agreement between theory and experiment for 
inclusive single-jet observables improves significantly.  
Furthermore, the dijet mass
distribution and the dijet angular distribution are well described by a
next-to-leading-logarithm calculation. The pseudorapidity dependence of the
dijet cross section in the region where both jets have negative 
pseudorapidities
is not well described, however. This has been shown to be due to the lack of
parton showering in the next-to-leading-logarithm calculation 
by comparing to HERWIG results with and without parton showering.

The theoretical description of the data in the regions dominated by the
resolved component appears to be good. There is thus reason to be optimistic
that the jet photoproduction data will soon reach a point where they can be
used to constrain photon parton distributions in conjunction with data from two
photon collisions from $e^+e^-$ facilities. However, with the current data the
experimental and theoretical uncertainties are comparable to the variations due
to changing the photon parton distribution set. Furthermore, the 
contribution of the
gluon distribution in the photon is rather small over the kinematic region
currently covered. The contribution increases as the pseudorapidities of both
jets are increased. If the data region can be extended, then the sensitivity to
the gluon distribution can be increased. Alternatively, if the photon energy
can 
be increased while holding the jet $E_T$ fixed, then $x_{\gamma}$ will be
decreased, thereby increasing the gluon contribution. This would require a data
sample with larger values of $y=E_{\gamma}/E_e$ or higher beam energies.

{\bf Note added.}  After completion of this work, we received \cite{kkk} 
which contains additional comparisions with ZEUS data as well as new 
comparisions with data from H1.

%
%----------------------------Journal macros---------------------------------
%
\def\Journal#1#2#3#4{{#1} {\bf #2}, #3 (#4)}
\def\CPC{\em Comp. Phys. Comm.}
\def\JCP{\em J. Comp. Phys.}
\def\JPG{{\em J. Phys.} G}
\def\NCA{\em Nuovo Cimento}
\def\NIM{\em Nucl. Instrum. Methods}
\def\NIMA{{\em Nucl. Instrum. Methods} A}
\def\NPB{{\em Nucl. Phys.} B}
\def\PLB{{\em Phys. Lett.}  B}
\def\PRL{\em Phys. Rev. Lett.}
\def\PRD{{\em Phys. Rev.} D}
\def\PR{\em Phys. Rev.}
\def\ZPC{{\em Z. Phys.} C}
\def\ZP{\em Z. Phys.}
%
%----------------------------References-------------------------------------
%

\newpage
\centerline{\bf \large{Figure Captions}}
\begin{description}
\item[Fig.\ 1]
The single-jet inclusive cross section as a function of $\eta_{\rm jet}$
integrated over $E_T^{\rm jet}>E_T^{\rm min}$ for 
$E_T^{\rm min} = 14,17,21$, and $25\,$GeV as measured by ZEUS 
\cite{zeuscone} using $R=1$ compared with our next-to-leading
order result for $R=1$ with $R_{\rm sep}=2R$ (top), $R_{\rm sep}=1.3R$ 
(middle), $R_{\rm sep}=R$ (bottom).
\item[Fig.\ 2]
The single-jet inclusive cross section as a function of $\eta_{\rm jet}$
integrated over $E_T^{\rm jet}>E_T^{\rm min}$ for 
$E_T^{\rm min} = 14,17,21$, and $25\,$GeV as measured by ZEUS 
\cite{zeuscone} using $R=0.7$ compared with our next-to-leading
order result for $R=0.7$ with $R_{\rm sep}=2R$ (top), $R_{\rm sep}=1.3R$ 
(middle), $R_{\rm sep}=R$ (bottom).
\item[Fig.\ 3]
The dijet inclusive cross section as a function of $ | \cos\theta^* | $ 
integrated over $M_{\rm JJ} > 47\,$GeV, and $M_{\rm JJ}$ integrated 
over $|\cos\theta^*|<0.8$ as measured by ZEUS \cite{zeusdijet} 
compared with our next-to-leading order result.
\item[Fig.\ 4]
The dijet inclusive cross section as a function of $E_T$ in bins of 
$\eta_1$ and $\eta_2$ as measured by ZEUS \cite{zeusdijet} 
compared with our next-to-leading order result.
\item[Fig.\ 5]
The dijet inclusive cross section as a function of $\eta_2$ in bins of 
$\eta_1$ integrated over $E_T>14\,$GeV as measured by ZEUS 
\cite{zeusdijet} compared with our next-to-leading order result 
for various renormalization-factorization scale choices.
\item[Fig.\ 6]
The dijet inclusive cross section as a function of $\eta_2$ in bins 
of $\eta_1$ integrated over $E_T>14\,$GeV as measured by ZEUS 
\cite{zeusdijet} compared with our next-to-leading order result 
for various photon parton distribution sets. 
\item[Fig.\ 7]
The dijet inclusive cross section as a function of $\eta_2$ in bins of 
$\eta_1$ integrated over $E_T>14\,$GeV for all $x_{\gamma}$ 
as measured by ZEUS 
\cite{zeusdijet} compared with our next-to-leading order result.  
The curves shown correspond to the full result (solid line), the 
resolved component with only quarks in the photon (medium dashed line), 
resolved component with only gluons in the photon (dotted line), and 
direct component (long dashed line).
\item[Fig.\ 8]
The dijet inclusive cross section as a function of $\eta_2$ in bins of 
$\eta_1$ integrated over $E_T>14\,$GeV for $x_{\gamma} > 0.75$ 
as measured by ZEUS 
\cite{zeusdijet} compared with our next-to-leading order result.  
The curves shown correspond to the full result (solid line), the 
resolved component with only quarks in the photon (medium dashed line), 
and direct component (long dashed line).
\item[Fig.\ 9]
The dijet inclusive cross section as a function of $\eta_2$ in bins of 
$\eta_1$ integrated over $E_T>14\,$GeV as measured by ZEUS 
\cite{zeusdijet} compared with 1.6 times the HERWIG \cite{herwig} result.
\item[Fig.\ 10]
The ratio of the HERWIG results to the HERWIG results at the leading order 
matrix element level,  {\it i.e.}
treating the two scattered partons as being the jets with
all showering and subsequent hadronization off, 
for the dijet inclusive cross section as a function 
of $\eta_2$ in bins of $\eta_1$.
\item[Fig.\ 11]
The ratio of the full next-to-leading order results to those obtained 
using only the leading order matrix elements 
for the dijet inclusive cross section 
as a function of $\eta_2$ in bins of $\eta_1$.
\end{description}

\newpage
%
%-------------------- figures -------------------------------------
%

% Fig. 1
\begin{figure}
\centerline{\hbox{\psfig{figure=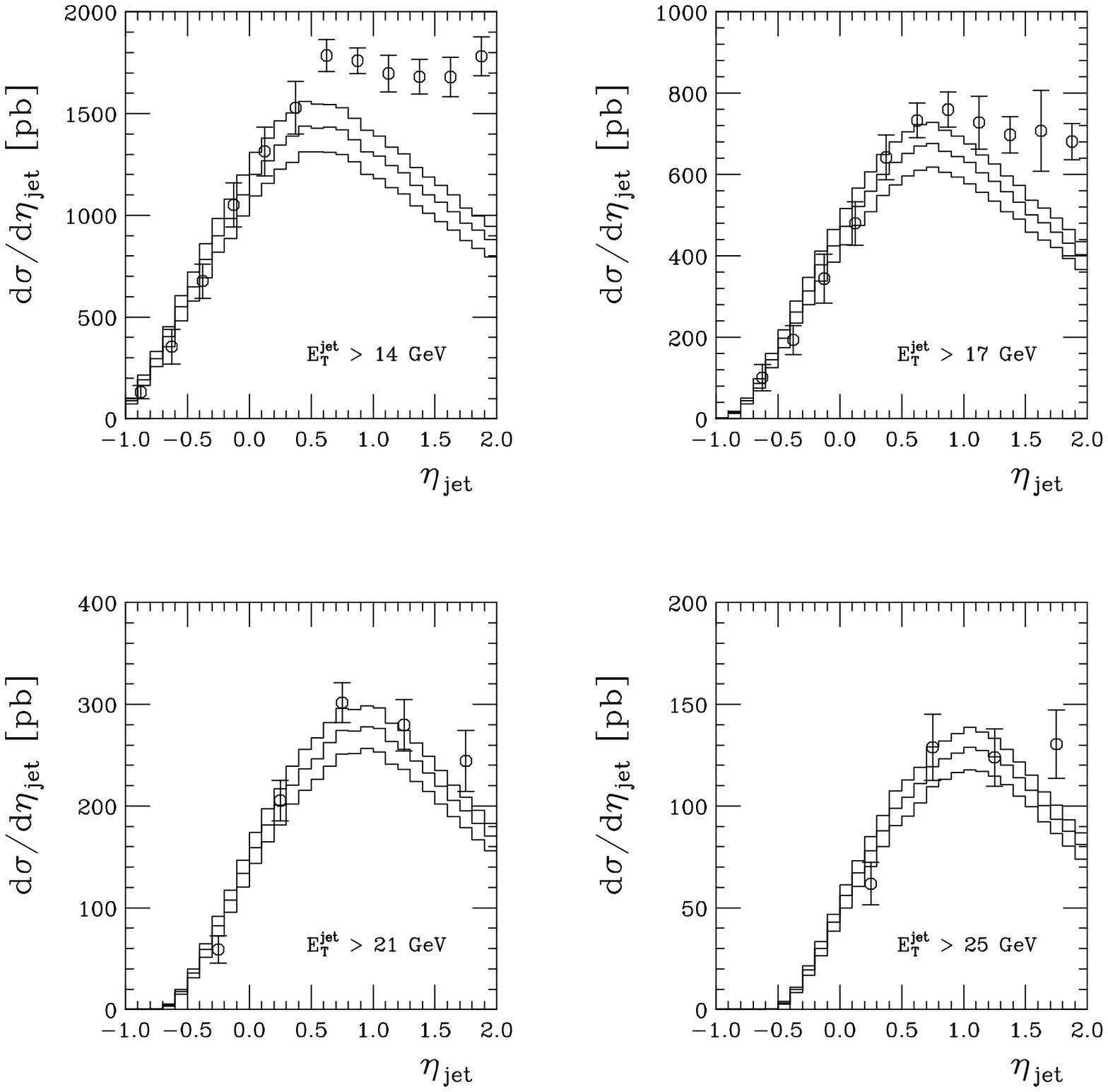,width=6.0in,height=4.2in}}}
\caption{}
\end{figure}

% Fig. 2
\begin{figure}
\centerline{\hbox{\psfig{figure=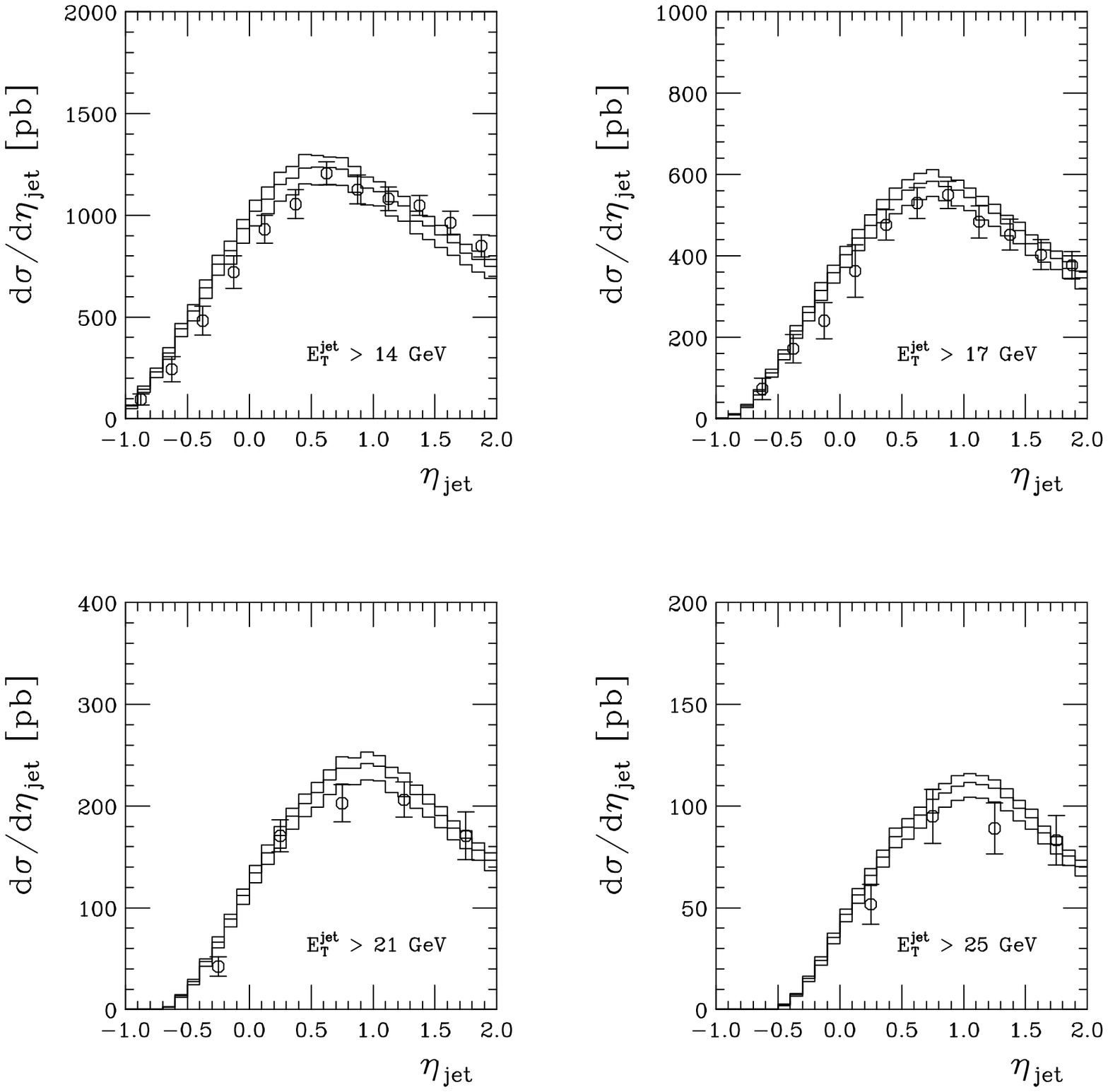,width=6.0in,height=4.2in}}}
\caption{}
\end{figure}

% Fig. 3
\begin{figure}
\centerline{\hbox{\psfig{figure=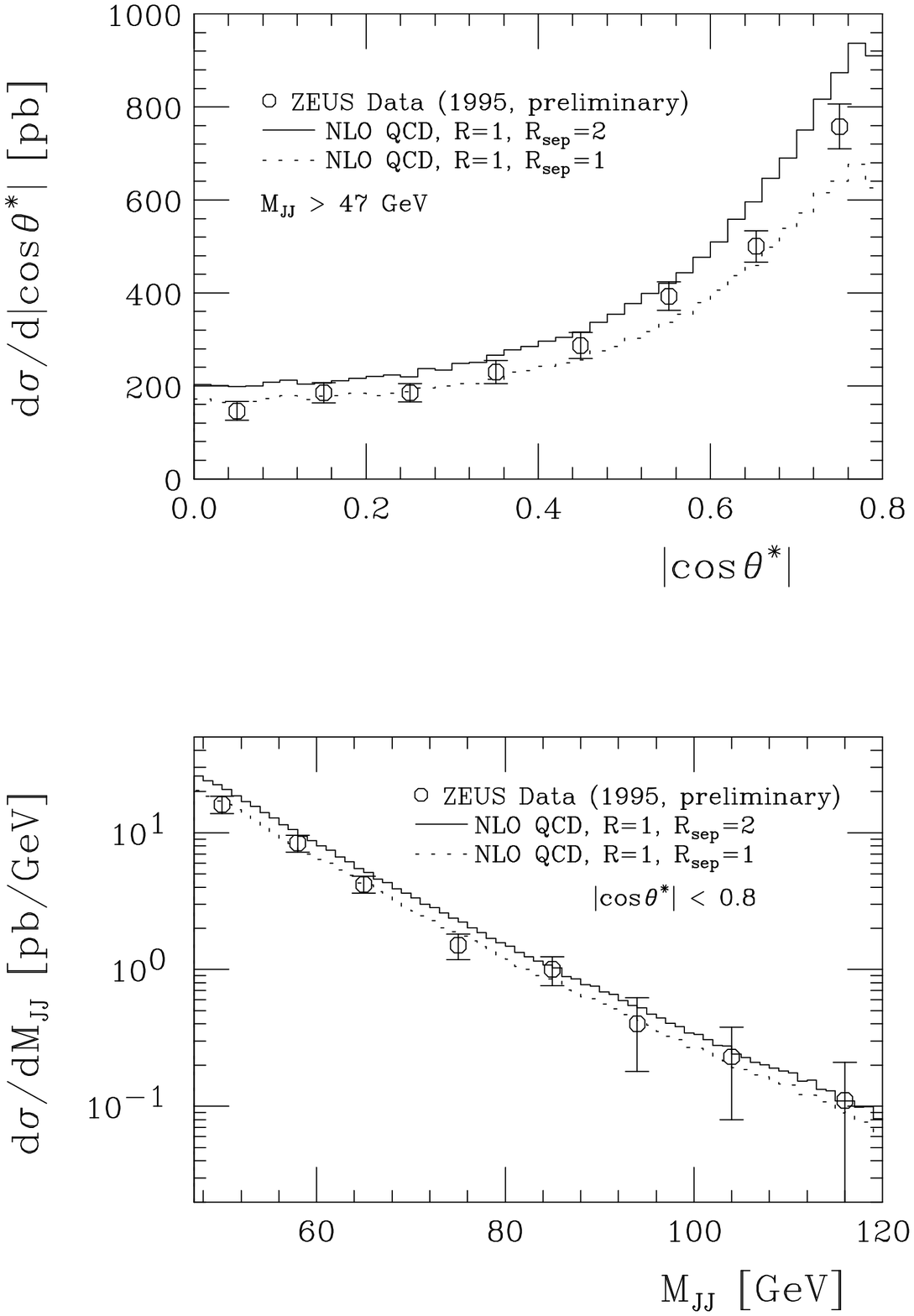,width=4.5in,height=5.0in}}}
\caption{}
\end{figure}

% Fig. 4
\begin{figure}
\centerline{\hbox{\psfig{figure=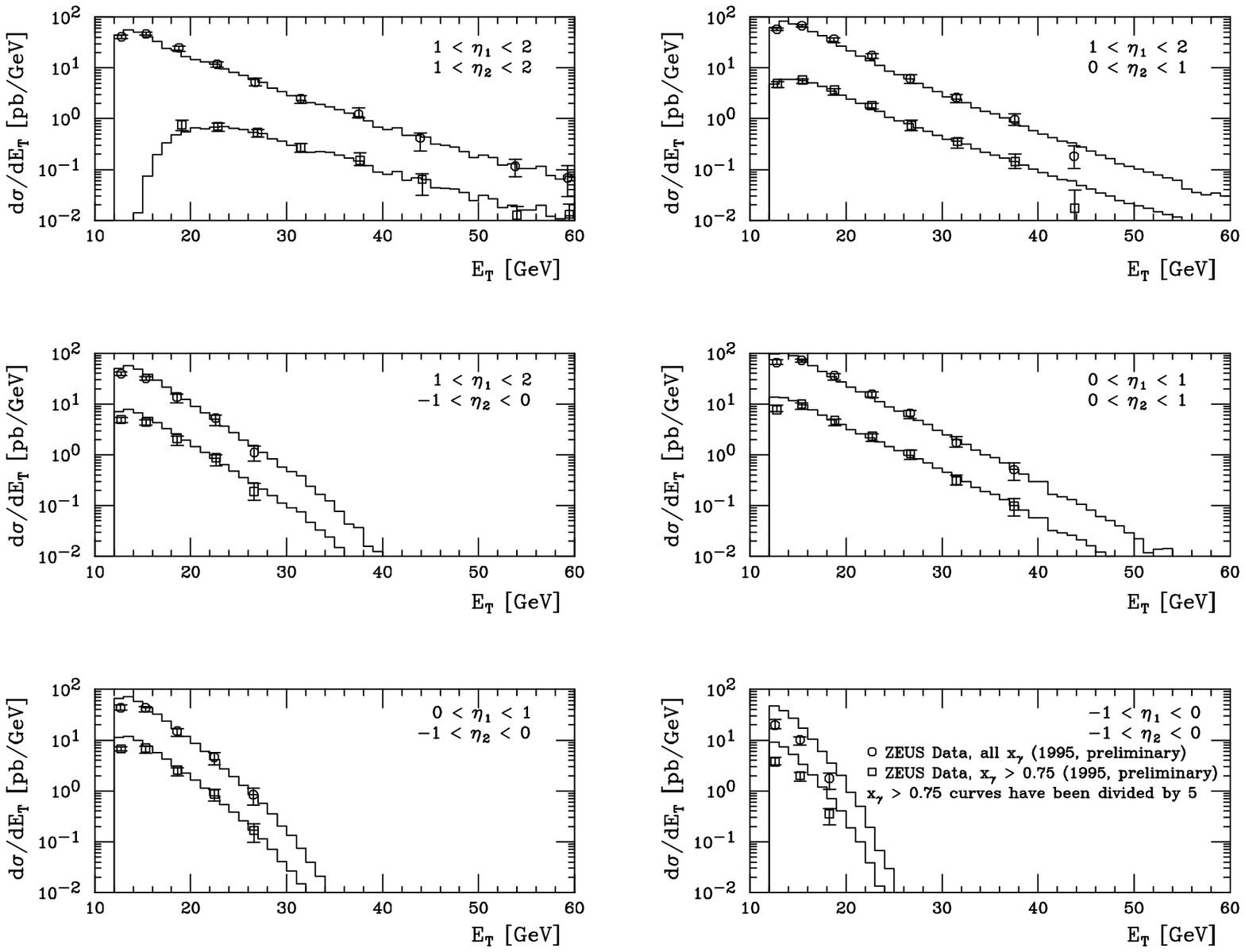,width=5.0in,height=6.0in}}}
\caption{}
\end{figure}

% Fig. 5
\begin{figure}
\centerline{\hbox{\psfig{figure=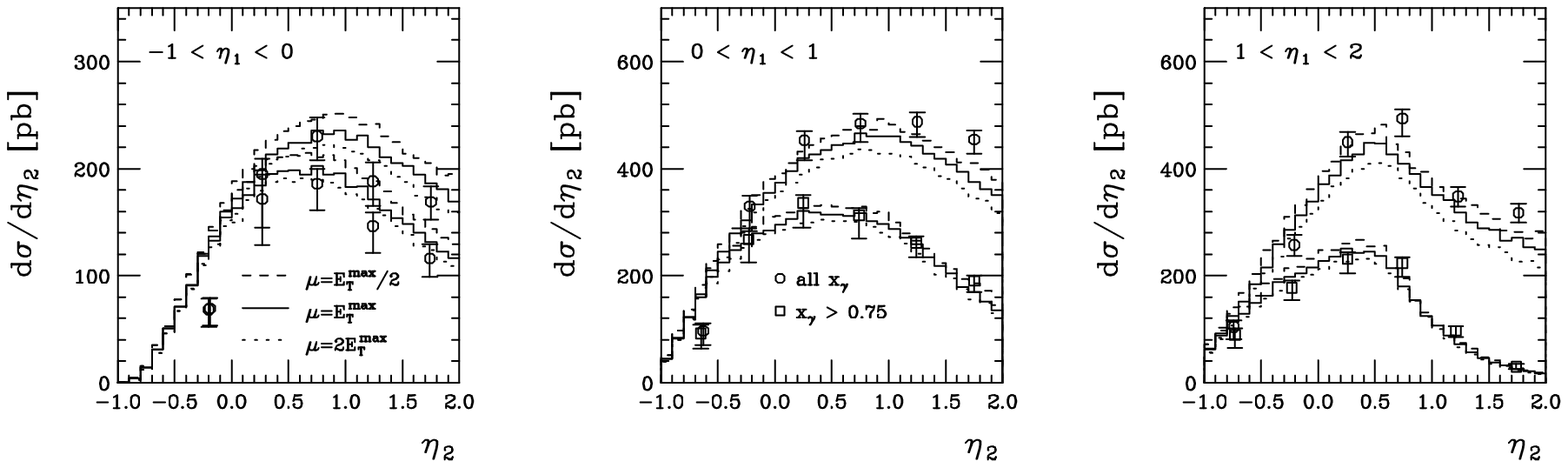,width=6in,height=2.5in}}}
\caption{}
\end{figure}

% Fig. 6
\begin{figure}
\centerline{\hbox{\psfig{figure=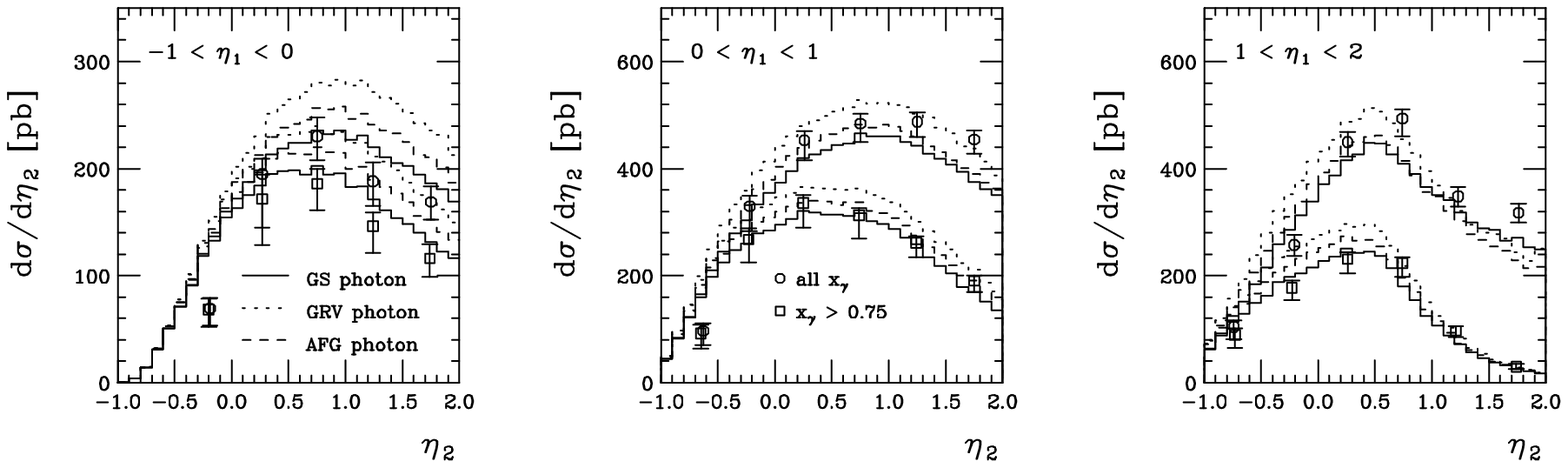,width=6in,height=2.5in}}}
\caption{}
\end{figure}

% Fig. 7
\begin{figure}
\centerline{\hbox{\psfig{figure=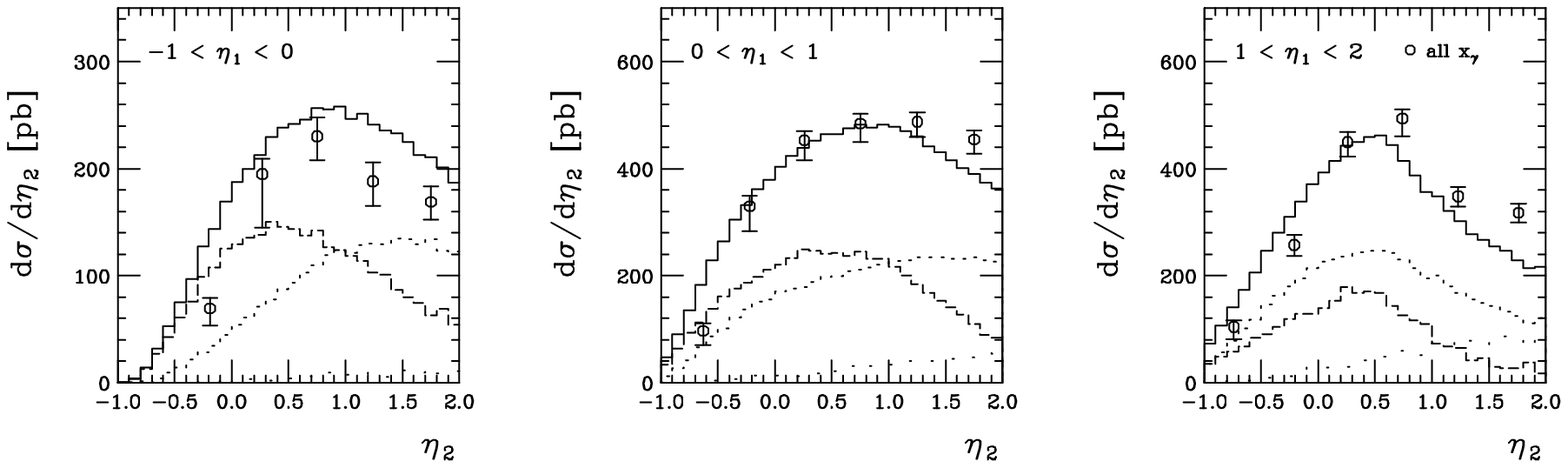,width=6in,height=2.5in}}}
\caption{}
\end{figure}

% Fig. 8
\begin{figure}
\centerline{\hbox{\psfig{figure=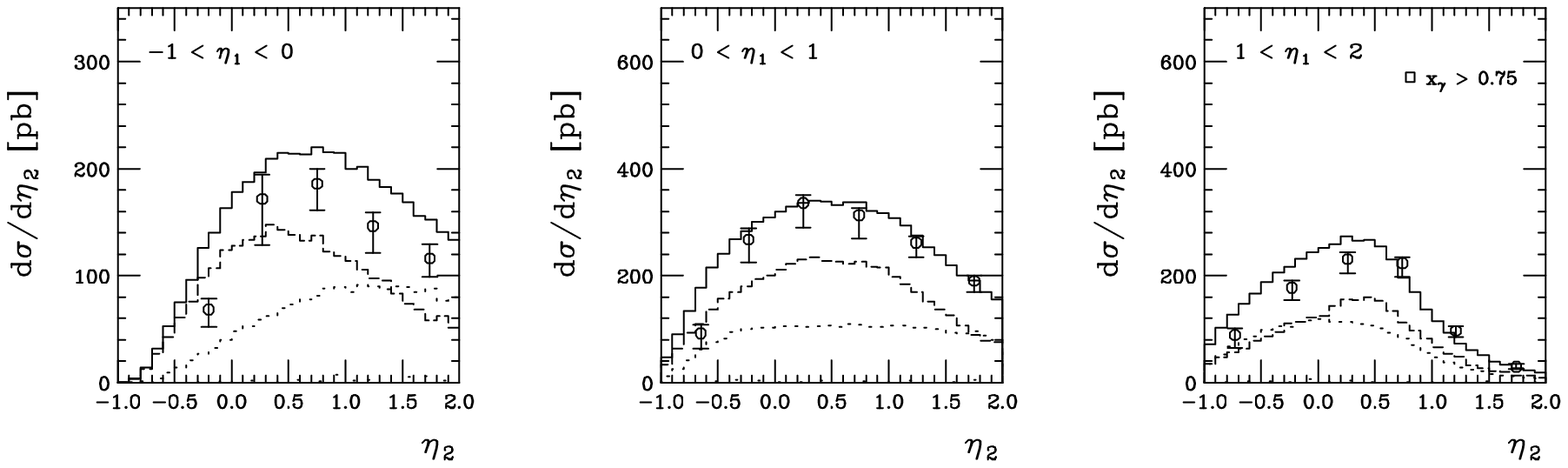,width=6in,height=2.5in}}}
\caption{}
\end{figure}

% Fig. 9
\begin{figure}
\centerline{\hbox{\psfig{figure=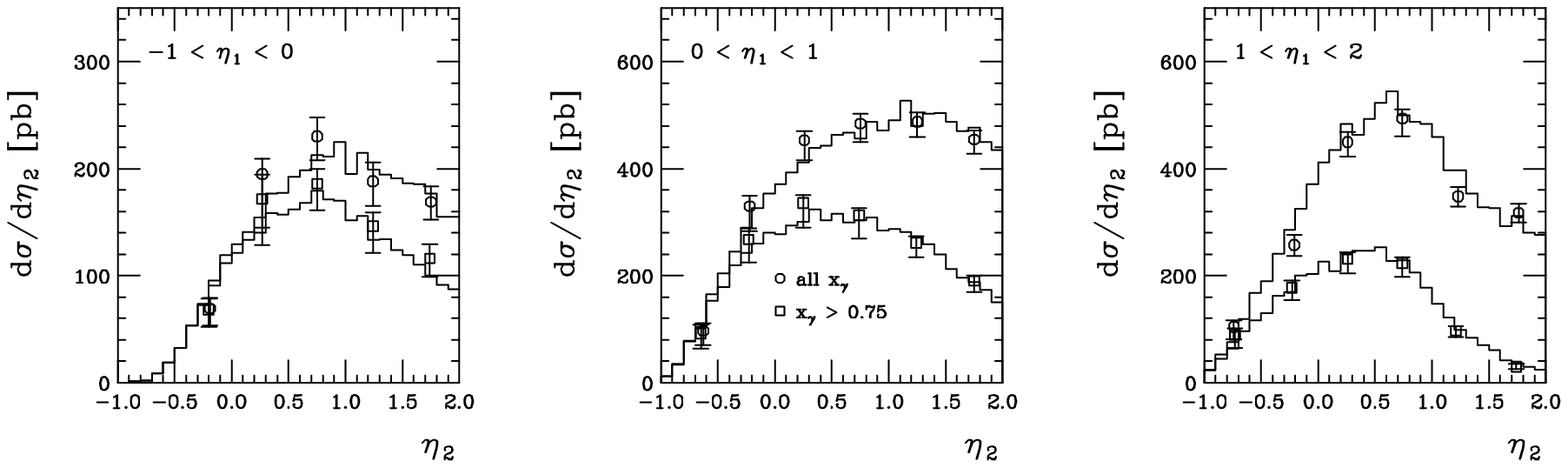,width=6in,height=2.5in}}}
\caption{}
\end{figure}

% Fig. 10
\begin{figure}
\centerline{\hbox{\psfig{figure=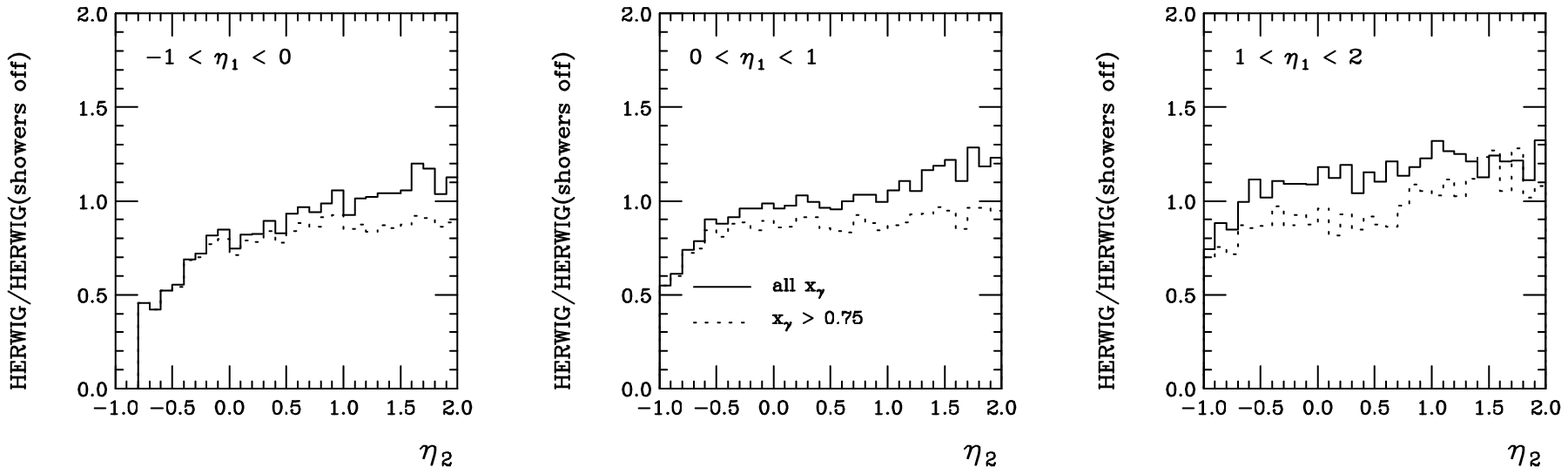,width=6in,height=2.5in}}}
\caption{}
\end{figure}

% Fig. 11
\begin{figure}
\centerline{\hbox{\psfig{figure=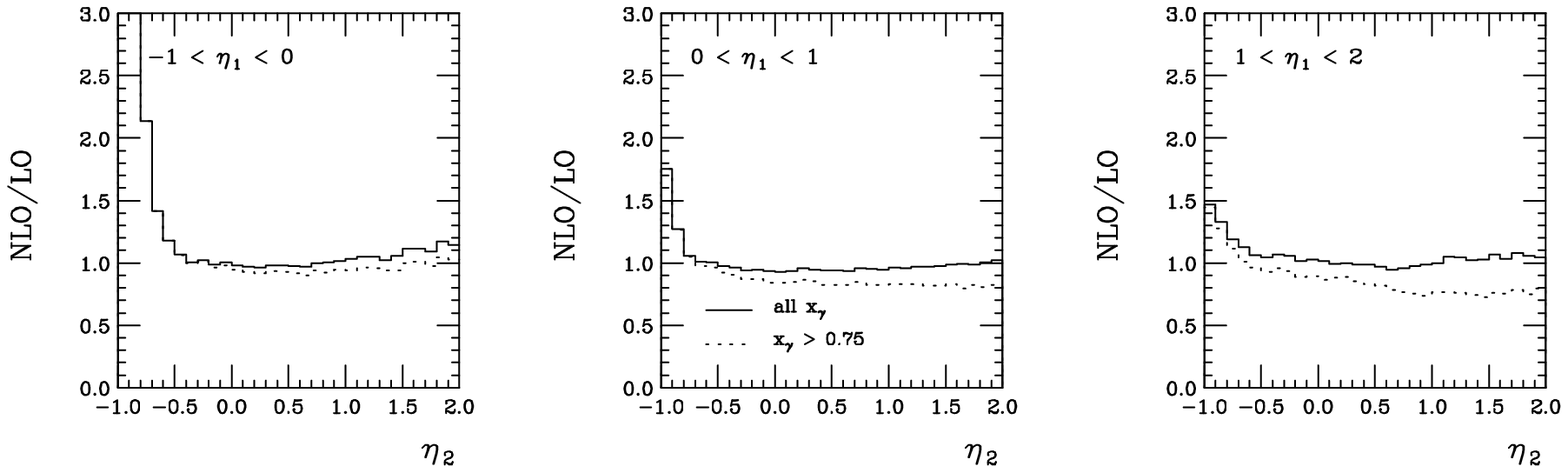,width=6in,height=2.5in}}}
\caption{}
\end{figure}

\end{document}